\documentclass[a4paper,USenglish,cleveref,autoref,thm-restate]{article}%{lipics-v2021}
\usepackage[dvipsnames,usenames]{xcolor}

\usepackage{amsthm,amssymb,amsmath}  
\usepackage{xspace,enumerate}
\usepackage[utf8]{inputenc}
\usepackage{thmtools}
\usepackage{blkarray}
\usepackage{chngcntr}
\RequirePackage{hyperref}
\hypersetup{colorlinks=true, urlcolor=Blue, citecolor=Green, linkcolor=BrickRed, breaklinks, unicode}

\renewcommand{\theenumii}{\theenumi \arabic{enumii}.}

\usepackage[ruled]{algorithm}
\usepackage[noend]{algpseudocode}
\algnewcommand{\LineComment}[1]{\State \(\triangleright\) #1}

\usepackage{graphicx}
\usepackage[noadjust]{cite}
  \usepackage{microtype}
    \usepackage{xspace}
    \usepackage{amsfonts}
    \usepackage{wasysym}
  \usepackage{makecell}
  \usepackage{tabularx}
  \usepackage{comment}
  \usepackage{todonotes}
  \usepackage{thmtools}
  \usepackage{verbatim}
\usepackage{units}
\usepackage{color}
\definecolor{darkblue}{rgb}{0,0.08,0.45}
  \usepackage{epsfig}
    \usepackage{graphics}
  \usepackage{graphicx}

\usepackage{marvosym}

\usepackage{booktabs}
\usepackage{etoolbox}
\usepackage{fullpage}

\newcommand{\shortproblemTwo}{Episode Matching}

\theoremstyle{plain}
\newtheorem{theorem}{Theorem}
\newtheorem{lemma}{Lemma}

\newtheorem{claim}{Claim}

\newtheorem{conjecture}{Conjecture}
\theoremstyle{definition}
\newtheorem{definition}{Definition}

\title{The Fine-Grained Complexity of Episode Matching}

\author{
    Philip Bille\\\texttt{phbi@dtu.dk} \and
    Inge Li Gørtz\\\texttt{inge@dtu.dk} \and
    Shay Mozes\\\texttt{smozes@idc.ac.il} \and
    Teresa Anna Steiner\\\texttt{terst@dtu.dk} \and 
    Oren Weimann\\\texttt{oren@cs.haifa.ac.il}
}

\begin{document}

\maketitle

\begin{abstract}
Given two strings $S$ and $P$, the \emph{Episode Matching} problem is to find the shortest substring of $S$ that contains $P$ as a subsequence. The best known upper bound for this problem is $\tilde O(nm)$~by Das et al. (1997)
, where $n,m$ are the lengths of $S$ and $P$, respectively. Although the problem is well studied and has many applications in data mining, this bound has never been improved. In this paper we show why this is the case by proving that no $O((nm)^{1-\epsilon})$ algorithm (even for binary strings) exists, unless the \emph{Strong Exponential Time Hypothesis (SETH)} is false.

We then consider the indexing version of the problem, where $S$ is preprocessed into a data structure for answering episode matching queries $P$. We show that for any $\tau$, there is a data structure using   $O(n+\left(\frac{n}{\tau}\right)^k)$ space that answers episode matching queries for any $P$ of length $k$ in  $O(k\cdot \tau \cdot \log \log n )$ time. We complement this upper bound with an almost matching lower bound, showing that any data structure that answers episode matching queries for patterns of length $k$ in time $O(n^\delta)$, must use $\Omega(n^{k-k\delta-o(1)})$ space, unless the {\em Strong $k$-Set Disjointness Conjecture}  is false. 
Finally, for the special case of $k=2$, we present a faster construction of the data structure using fast min-plus multiplication of bounded integer matrices. 
\end{abstract}

\section{Introduction}
 A string $P$ is a \emph{subsequence} of a string $S$ if $P$ can be obtained by deleting characters from $S$. Given two strings $S$ and $P$, the \emph{Episode Matching} problem~\cite{cpm/DasFGGK97}  is to find the shortest substring of $S$ that contains $P$ as a subsequence. In its indexing version, we are given $S$ in advance and we need to preprocess it into a data structure for answering episode matching queries $P$.

 \subsection{Episode matching}
 The Episode Matching problem was introduced by Das et al.~\cite{cpm/DasFGGK97} in 1997 as a simplified version of the frequent episode discovery problem first studied by Mannila et al.~\cite{datamine/MannilaTV97}. Das et al.~\cite{cpm/DasFGGK97} gave an upper bound of $O(nm/\log n)$, where $n$ is the length of $S$ and $m$ is the length of $P$. 
 Even though the problem and its variations have been thoroughly studied \cite{iandc/ApostolicoA02,fuin/KuriNM03,apal/BoassonCGM01,ipl/CegielskiGM06,jal/MakinenNU05,njc/CrochemoreIMRTT02} and have many applications in data mining~\cite{jnsm/KlemettinenMT99,icdm/AtallahGS04,kais/GwaderaAS05,datamine/MannilaTV97,iandc/ApostolicoA02}, the original $O(nm/\log n)$ upper bound has never been improved. 
     In Section~\ref{sec:lower_bound} we show why this is the case, by proving a lower bound conditioned on the \emph{Strong Exponential Time Hypothesis (SETH)} (actually, on the Orthogonal Vectors Hypothesis (OV)). This continues a recent line of research on quadratic lower bounds for string problems conditioned on SETH \cite{focs/BringmannK15,focs/BackursI16,icalp/EquiGMT19,algorithmica/KociumakaRS19,ipl/Polak18,algorithmica/DurajKP19,spire/Gibney20,jcomp/BackursI18,focs/AbboudBW15,icalp/AbboudWW14}. Our reduction is very simple and easily extends to binary alphabet and to unbalanced inputs.
   Our result is summarized by the following theorem. 
     \begin{theorem}\label{thm:lower_bound_ovh} For any $\epsilon>0$, Episode Matching on binary strings of lengths $n$ and $m = n^{\alpha}$ (for any fixed $\alpha\le 1$) cannot be solved in $O((mn)^{1-\epsilon})$ time, unless SETH is false.
    \end{theorem}
\medskip
\noindent
{\bf Related work.} A related problem to episode matching is the \emph{Longest Common Subsequence (LCS)} problem. In that problem, the goal is to find for two strings $S$ and $P$ (of lengths $n$ and $m$ respectively) the longest string that is a subsequence of both $S$ and $P$. In 2015 Abboud et al. \cite{focs/AbboudBW15} and Bringmann and Künnemann \cite{focs/BringmannK15} independently proved quadratic lower bounds for LCS conditioned on the SETH. That is, they showed that assuming SETH, there cannot be an $O(m^{2-\epsilon})$ algorithm. However, there exists an $\tilde{O}(n+m^2)$ algorithm for LCS \cite{jacm/Hirschberg77}. This suggests that for unbalanced inputs where $m\ll n$, episode matching is harder than LCS. 
    
    Further, the episode matching problem can be seen as a special case of the \emph{Approximate String Matching} problem. There, given strings $S$ and $P$, the goal is to find the substring of $S$ minimizing some distance measure to $P$. If this distance measure is the number of deleted characters from $S$, the problem is equivalent to episode matching. Another version of the approximate string matching problem uses \emph{edit distance}. Backurs and Indyk \cite{jcomp/BackursI18} implicitly give a quadratic lower bound conditioned on SETH for that version as a stepping stone for achieving a lower bound for edit distance, using an alphabet of size seven. This bound does not however directly translate to our problem, and uses a more complicated construction and a larger alphabet.
    
To avoid misunderstandings, we note that in the paper by Mannila et al.~\cite{datamine/MannilaTV97}, an \emph{episode} was more generally defined as a collection of events that occur together. The string formulation by Das et al.~\cite{cpm/DasFGGK97} is a simplification of this original definition, thus, the terms \emph{episode} and \emph{episode matching} have been used to name different concepts in the literature, see e.g.  \cite{dke/AcharAS13, dasfaa/TagoAKMI12,sdm/GwaderaAS05,dis/HiraoISTA01,nowickacomplexity}. We only consider the string formulation by Das et al.~\cite{cpm/DasFGGK97}.

 \subsection{Episode Matching Indexing} 
 Limited by the above lower bound for episode matching, a natural question to ask is what bounds we can get if we allow preprocessing. Alas, in terms of time complexity, preprocessing does not help. Namely, our reduction is such that each of the two sets in the Orthogonal Vectors instance (see Definition~\ref{def:orthogonal}) is encoded {\em independently} into one of the two strings in the episode matching instance.  This implies (see e.g.~\cite{EquiMT21}) that episode matching cannot be solved in $O((mn)^{1-\epsilon})$ time even if we are allowed polynomial time to preprocess one of the two strings in advance. In other words, polynomial time preprocessing is not enough to guarantee subquadratic time queries. In fact, this holds even when one of the strings is fixed (cf.~\cite[Section 1.1.2]{AbboudW21}). We therefore focus on time-space tradeoffs.  

Formally, given a string $S$ and an integer $k$, the {\em episode matching indexing} problem asks to construct a compact data structure that can quickly report the episode matching of any pattern $P$ of length $k$ (i.e. compute the length of the shortest substring of $S$ that contains $P$ as a subsequence). 
Apostolico and Atallah \cite{iandc/ApostolicoA02} gave a linear (optimal) space data structure whose query time may be prohibitive (depending on the number of distinct minimal substrings containing a prefix of the pattern as a subsequence). In Section~\ref{subsec:tstradeoff} we present a time-space tradeoff with faster query time: 
 
 \begin{theorem}\label{thm:ds_upper_bound}
For any $\tau$, there is a data structure using space $O(n+\left(\frac{n}{\tau}\right)^k)$ that answers Episode Matching queries in time $O(k\cdot \tau \cdot \log \log n )$.
\end{theorem}

In Section~\ref{sec:lowerindexing} we give the following almost matching lower bound, conditioned on the $k$-Set Disjointness Conjecture:

\begin{theorem}\label{thm:ds_lower_bound}
For any $\delta\in[0,1/2]$, a data structure that answers Episode Matching queries in time $O(n^\delta)$ must use $\Omega(n^{k-k\delta-o(1)})$ space, unless the Strong $k$-Set Disjointness Conjecture is false.
\end{theorem}

Finally, in Section~\ref{sec:kequals2} we consider the decision version of the case $k=2$. That is,  we are given some threshold $t$, and a query $(a,b)$ need only report whether $S$ contains a substring of length at most $t$ that starts with the letter $a$ and ends with the letter $b$. 
In this setting, there is a simple $O(1)$-query time $O(\sigma^2)$-space data structure for episode matching (where $\sigma$ is the size of the alphabet): precompute and store the $\sigma \times \sigma$  matrix $D$ with the answers to every possible query. Naively, we can compute $D$ in time $\min\{\tilde{O}(n\sigma+\sigma^2),O(nt+\sigma^2)\}$. We show that for various values of $\sigma$ and $t$, we can compute the matrix $D$ faster by using fast min-plus matrix multiplication of bounded integer matrices~\cite{DBLP:conf/soda/WilliamsX20}:  

 \begin{theorem}\label{thm:minplus}
 For a given threshold $t$, we can compute the matrix $D$ in time
 
$O(\sigma^{1/2+\omega/2}\left(\frac{n}{t}\right)^{1/2} \sqrt{n})$ if $\sigma<\frac{n}{t}$, and 

   $O(\sigma^2\left(\frac{n}{t}\right)^{\omega/2-1}\sqrt{n})$ if $\sigma\ge \frac{n}{t}$.
\end{theorem}

In particular, using the current matrix multiplication exponent $\omega<2.4$ \cite{soda/AlmanW21}, if $\sigma<\frac{n}{t}$ then we get $O(\sigma^{1.7}nt^{-1/2})$ (which is smaller than $\min(n\sigma,nt)$ for any $t > \sigma^{1.4}$), and if $\sigma\ge \frac{n}{t}$ we get $O(\sigma^{2.2}\sqrt{n})$ (which is smaller than $O(n\sigma)$ for any $\sigma < n^{0.416}$).

\section{Preliminaries}
In this section we will review some basic string notation and important hypotheses. 

A string~$S$ of length~$n$ is a sequence $S[0]\cdots S[n-1]$ of $n$ characters drawn from an alphabet $\Sigma$ of size $|\Sigma|=\sigma$. A string $S[i\cdots j]=S[i]\cdots S[j]$ for $0\leq i < j < n$ is called a \emph{substring} of $S$. For two strings $X$ and $Y$ we denote their \emph{concatenation} as $X\cdot Y$ or $XY$.

The Strong Exponential Time Hypothesis is a popular conjecture about the hardness of the $k$-SAT problem, postulated by Impagliazzo and Paturi \cite{jcss/ImpagliazzoP01}. The $k$-SAT problem is to decide whether there exists a satisfying assignment for a Boolean formula on $n$ variables and clauses of width at most $k$.
    \begin{conjecture}[The Strong Exponential Time Hypothesis] \label{con:SETH} There is no $\epsilon>0$ for which $k$-SAT can be solved in time $O(2^{(1-\epsilon)n})$ for all $k\geq 3$. \end{conjecture}

Instead of reducing directly from $k$-SAT, we reduce from the Orthogonal Vectors problem, and use two conjectures about the hardness of Orthogonal Vectors which are implied by SETH for $d=\omega(\log n)$.

\begin{definition}[Orthogonal Vectors Problem (OV)] Given two sets $A=\{a_1,\dots, a_n\}$ and $B=\{b_1,\dots, b_m\}$ of $d$-dimensional binary vectors, decide whether there is an orthogonal pair of vectors $a_i\in A$ and $b_j\in B$. \label{def:orthogonal}
\end{definition}
The two conjectures consider the cases of equal set size and unbalances set size, respectively. They roughly state that any algorithm solving OV that has a polynomial dependency on the dimension (denoted poly$(d)$), cannot achieve a significantly better asymptotic running time than the product of the two set sizes.

    \begin{conjecture}[Orthogonal Vectors Hypothesis (OVH)] For $|A|=|B|=n$, there is no $\epsilon>0$ for which OV can be solved in time $O(n^{2-\epsilon}\mathrm{poly}(d))$.
    \end{conjecture}
    
    \begin{conjecture}[Unbalanced Orthogonal Vectors Hypothesis (UOVH)] Let $0<\alpha\leq 1$, $|A|=n$ and $|B|=m$. There is no $\epsilon>0$ for which OV restricted to $m=\Theta(n^{\alpha})$ and $d= n^{o(1)}$ can be solved in time $O((nm)^{1-\epsilon})$.\end{conjecture}

\noindent   It is known that SETH implies both OVH and UOVH~\cite{tcs/Williams05,focs/BringmannK15}. Finally, we consider the following conjecture of Goldstein et al.~\cite{DBLP:conf/wads/GoldsteinKLP17} on the $k$-Set Disjointness problem:
 
\begin{definition}[$k$-Set Disjointness Problem] Preprocess $m$ sets $S_1$, $S_2$, \dots, $S_m$ of total size $\sum_{i=1}^m |S_i|=N$ drawn from a universe $U$ such that given $(i_1,i_2,\dots, i_k)$ we can quickly decide whether $\bigcap_{j=1}^k S_{i_j}=\emptyset$.
\end{definition}
\begin{conjecture}[Strong $k$-Set Disjointness Conjecture]\label{con:kset} Any data structure for the $k$-Set Disjointness Problem that answers queries in time $T$ must use $\tilde{\Omega}\left(N^k/T^k\right)$ space.
\end{conjecture}

\section{Episode Matching} \label{sec:lower_bound}

In this section we prove Theorem \ref{thm:lower_bound_ovh}. We first prove it for an alphabet of size four, and then for a binary alphabet.% \todo{what does $m\ll n$ mean?}

\subsection{Alphabet of Size Four}\label{sec:sizefour}
We show how to reduce an instance of OV to Episode Matching with alphabet $\{0,1,x,\$\}$. Let $A=\{a_1,\dots, a_n\}$ and $B=\{b_1,\dots, b_m\}$ be two sets of vectors in $\{0,1\}^d$. Without loss of generality, we assume $m\leq n$. %Let $|A|=n$ and $|B|=m\leq n$.%=\Theta(n^{\alpha})$. 
We show how to construct a string $P$ of length $2dm+1$ and a string $S$ of length $3d(4n+1)+1$ such that there is a pair of orthogonal vectors $a_i\in A$ and $b_j\in B$ if and only if there is a substring of $S$ of length at most $3d(2m-1) +1$ that contains $P$ as a subsequence. 
    
\medskip
\noindent
{\bf Constructing $P$.}
We construct $P$ from the set $B$ in the following way. For every $b\in B$, we define $p(b)$ as the string of length $2d-1$ obtained by inserting an $x$ symbol between each consecutive entries of $b$. That is, 
 $$	
p(b) = b[0]\  x\  b[1]\  x\ \cdots x\  b[d]
 $$
Then, $P$ is a string of length $2dm+1$ defined as the concatenation: 
 $$
 	P = \$\  p(b_1)\ \$\  p(b_2)\ \$\cdots\$\  p(b_m)\ \$
 $$

\medskip
\noindent
{\bf Constructing $S$.}
We construct $S$ from the set $A$. For a vector $a$ in $\{0,1\}^d$, we define for each entry in $a$ a string of length 2, called a \emph{coordinate gadget}:
    $$
    	s(a[i])=01\text{ if } a[i]=0, $$$$ s(a[i])=00 \text{ if } a[i]=1.
    $$
    Then, $s(a)$ is a string of length $3d-1$ 
    defined as the concatenation: 
     $$
     	s(a) = s(a[0])\  x\  s(a[1])\  x\cdots x\  s(a[d])
     $$
      For example, the vector 
    $a=10001$ defines the string $s(a)=00x01x01x01x00$.
Let $z$ be the $d-$dimensional zero vector, and so  $s(z)=01x01x\dots x01$. Then, $S$ 
  consists of $\$\  s(z)\ \$$ followed by two copies of the concatenation of $s(a_i)\ \$\  s(z)\ \$$ for $1\leq i \leq n$. That is:
               $$S=\$ s(z)\$ s(a_1)\$ s(z)  \$ s(a_2)  \$\cdots\$ s(z) \$  s(a_n)\$ s(z)\$ s(a_1)\$ s(z)  \$ s(a_2) \$\cdots\$ s(z) \$  s(a_n)\$  s(z) \$$$  
    
  \noindent We call a substring of $S$ of the form ``$s(a)\  \$$'' or ``$s(z)\  \$$'' a \emph{block}. A block has length $3d$.
  The string $S$ contains $2n$ blocks for the elements of $A$, plus $2n+1$ blocks for the separating $s(z)$, thus $4n+1$ blocks in total. There is an extra $\$$ at the start of the string. Thus, $|S|=3d(4n+1)+1=\Theta(d n)$.
\medskip
\noindent
{\bf Correctness.}
For two strings $Y$ and $X$, an \emph{alignment $L$ of $Y$ in $X$} is a sequence $0\leq j_0<\dots<j_{|Y|-1}\leq |X|-1$
such that $Y[\ell]=X[j_{\ell}]$ for every $0\leq \ell \leq |Y|-1$. We say  $Y[\ell]$ is \emph{aligned} to $X[j_\ell]$ and that $L$ \emph{spans} the substring $X[j_0 \cdots j_{|Y|-1}]$. 
Clearly, $Y$ is a subsequence of $X$ if and only if there exists an alignment of $Y$ in $X$.

The following Lemma establishes the translation of orthogonality in our reduction.

    \begin{lemma} \label{lem:subseq_four}  For two vectors $a$ and $b$ of dimension $d$, the string $p(b)$ is a subsequence of $s(a)$ if and only if $b$ and $a$ are orthogonal.
    \end{lemma}
    
    \begin{proof}
    
  If $a$ and $b$ are orthogonal, then for every $b[i]=1$ we have that $a[i]=0$ and therefore $s(a[i])=01$. Thus we can align $b[i]$ to the 1 in $s(a[i])$. If $b[i]=0$, we can align $b[i]$ to the first character in $s(a[i])$. We can therefore define an alignment where we align every $b[i]$ to one of the characters in $s(a[i])$, and align the $i$th $x$ in $p(b)$ to the $i$th $x$ in $s(a)$. Therefore $p(b)$ is a subsequence of $s(a)$.
  
  On the other hand, if $p(b)$ is a subsequence of $s(a)$, then since $p(b)$ and $s(a)$ both contain exactly $d-1$ $x$'s, any alignment of $p(b)$ in $s(a)$ \emph{must} align the $i$th $x$ in $p(b)$ to the $i$th $x$ in $s(a)$. Thus, each $b[i]$ must be aligned to a character in $s(a[i])$. 
    If $b[i]=1$, we can align it with a character in $s(a[i])$ only if $a[i]=0$.
    Thus, if $p(b)$ is a subsequence, then for every $b[i]=1$, we must have $a[i]=0$. Therefore $a$ and $b$ are orthogonal.
    \end{proof}
Lemma~\ref{lem:subseq_four} implies that any $p(b)$ is a subsequence of $s(z)$. 
This immediately yields a substring of $S$ of length $3d(2m-1)+1$ which contains $P$ as a subsequence:  If we align $P$ with $m$ consecutive occurrences of $s(z)$, 
   the resulting substring contains a $\$$ at the beginning, $m$ blocks of the form ``$s(z)\ \$$'' and $m-1$ blocks of the form ``$s(a)\ \$$''. Next, we show that if there is no pair of orthogonal vectors, then there is no shorter substring of $S$ that contains $P$ as a subsequence.
\begin{claim} If there exist no $a\in A$ and $b\in B$ which are orthogonal, then there exists no substring of $S$ of length $< 3d(2m-1)+1$ which contains $P$ as a subsequence.\end{claim}
 \begin{proof}
     Consider an alignment $L$ of $P$ in $S$. If for every $b\in B$ the substring $p(b)$ is aligned to a string of the form $s(z)$, then $L$ spans a string of length at least $3d(2m-1)+1$. Consider now the case where there exists a $b\in B$ such that $p(b)$ is not fully aligned to a string of the form $s(z)$. By Lemma~\ref{lem:subseq_four}, since there is no $a\in A$ such that $a$ and $b$ are orthogonal, there is no $s(a)$ such that $p(b)$ is fully aligned within $s(a)$. Since no $s(a)$ or $s(z)$ contain a $\$$, this means that the alignment of the string $\$~p(b)~\$$ spans a string containing either a substring of the form $\$~s(a)~\$~s(z)~\$$ or a substring of the form $\$~s(z)~\$~ s(a)~\$$. Then the alignment $L'$ defined by aligning $\$~p(b)~\$$ to the $\$~s(z)~\$$ in that substring, and everything else as in $L$, spans a substring in $S$ no longer than the substring spanned by $L$. Repeating this for each $p(b)$ that is not aligned to $s(z)$ gives an alignment where every $p(b)$ is aligned to some $s(z)$, and which spans a substring in $S$ no longer than the substring spanned by $L$. Since this substring contains at least $m$ copies of $\$~s(z)~\$ $, it has length at least $3d(2m-1)+1$. 
 \end{proof}
 Next, we show how to align $P$ in $S$ to yield a shorter substring, if there does exist an orthogonal pair.
\begin{claim}\label{claim_short} If there exist $a\in A$ and $b\in B$ which are orthogonal, then there is a substring of $S$ of length $< 3d(2m-1)+1$ that contains $P$ as a subsequence.\end{claim}
\begin{proof} Assume $a_i$ and $b_j$ are orthogonal and $j\leq i$. We align $p(b_j)$ to the first copy of $s(a_i)$, and align $p(b_{j+1})\cdots p(b_m)$ to the next $m-j$ copies of $s(z)$ to the right of $s(a_i)$, and $p(b_1) \cdots(b_{j-1})$ to the previous $j-1$ copies of $s(z)$ to the left of $s(a_i)$. See Figure \ref{fig:alignment}. This is possible since $j \leq i$.  Let  $T$ denote the resulting substring of $S$.

    \begin{itemize}
        \item Case 1: $j=1$ (or $j=m$). In this case, the substring of $T$ spans $2m-2$ blocks: It starts (ends) with ``$\$\  s(a_i)\ \$$'', and then includes the $m-1$ following (preceeding) $s(z)$ blocks. Between any two $s(z)$ blocks, there is another block corresponding to some $a_\ell\in A$. Thus in total, the length of $T$ is $3d(2m-2) + 1 < 3d(2m-1) +1$. 
        \item Case 2: $1<j<m$.
        The substring $T$ starts and ends with $\$~s(z)~\$$, and we align to $m-1$ of the $s(z)$ blocks and to $s(a_i)$ which is somewhere inbetween these. Thus, $T$ includes $m-1+m-2=2m-3$ blocks, and the length of $T$ is $3d(2m-3) +1 < 3d(2m-1) +1 $. %$(3d-1)(2n-3)+2n-2$
        \end{itemize}
      If $j>i$, we align $p(b_j)$ to the \emph{second} copy of $s(a_i)$, and again align  $p(b_{j+1})\cdots p(b_m)$ to the next $m-j$ copies of $s(z)$ to the right of $s(a_i)$, and $p(b_1)\cdots p(b_{j-1})$ to the preceding $j-1$ copies of $s(z)$ to the left of $s(a_i)$. This is possible since $m-j\leq n-i$. The rest follows analogously.
        \end{proof}
        
        \begin{figure}
   \centering
       \setlength{\tabcolsep}{0pt}
    \begin{tabular}{ccccccccc}
         $\cdots ~$ & $\$~\phantom{_{j2}}s(z)_{\phantom{-}}~\$$ & $~s(a_{i-1})$~ &$\$~\phantom{_{j1}}s(z)_{\phantom{-}}~\$$~ & $~s(a_i)~$& $\$~\phantom{_{j1}}s(z)_{\phantom{+}}$ & $\$~s(a_{i+1}) ~$& $\$~\phantom{_{j2}}s(z)_{\phantom{+}}~\$$ & $~\cdots$ \\
         $\cdots~$ & $\$~p(b_{j-2})~~~$&  & $\$~p(b_{j-1})~\$$~ & $~p(b_j)~$ & $\$~p(b_{j+1})$ & & $\$~p(b_{j+2})~ \$$ & $~\cdots$

    \end{tabular}
    \caption{The alignment of $P$ and $S$ as described in the proof of Claim \ref{claim_short}.}\label{fig:alignment}
    \end{figure}

\medskip
\noindent
{\bf Analysis.}
When $m=\Theta(n)$, we have that $|P|=\Theta(|S|) = \Theta(nd)$, which means that an $O((|S||P|)^{1-\epsilon})$ algorithm for \shortproblemTwo\ implies an $O(n^{2-2\epsilon}d^{2-2\epsilon})$ algorithm for OV, contradicting OVH.  
   
When $m=\Theta(n^{\alpha})$ for some $0 < \alpha < 1$, and $d= n^{o(1)}$%\todo{SM: $d= n^{o(1)}$}
, the length of $P$ is $2dm+1=mn^{o(1)}$, and the length of $S$ is $3d(4n+1)+1=n^{1+o(1)}$. This means that an $O((|S||P|)^{1-\epsilon})$ algorithm for \shortproblemTwo\ implies an $(nm)^{1-\epsilon}n^{o(1)}=O((nm)^{1-\epsilon'})$ algorithm for OV for any $\epsilon'<\epsilon$, contradicting UOVH.
  
  This proves Theorem \ref{thm:lower_bound_ovh} for alphabet size at least 4.

\subsection{Binary Alphabet}\label{sec:binary}
We now prove Theorem \ref{thm:lower_bound_ovh} for a binary alphabet. We will use almost the same reduction as before, but replace $x$ and $\$$ with binary gadgets.

\medskip
\noindent
{\bf Inner separator gadget.} Instead of the separating symbol $x$, we define an \emph{inner separator gadget} of the form $0^d$. 
The strings $p(b)$ and $s(a)$ are defined as before, except that each $x$ is substituted with the inner separator gadget $0^d$, and we add an extra inner separator gadget at the beginning and end of the string.  That is,  $$p(b) = 0^d\  b[1]\  0^d\  b[2]\  0^d\cdots 0^d\  b[d]\  0^d $$ and $$s(a) =  0^d\  s(a[1])\  0^d\  s(a[2])\  0^d \cdots 0^d\  s(a[d])\  0^d\;.$$

\medskip
\noindent
{\bf Outer separator gadget.} Instead of the separating symbol $\$$, we define an \emph{outer separator gadget} of the form $1^{d+1}$.
The strings $P$ and $S$ are defined as before, with the new versions of $p(b)$, $s(a)$ and $s(z)$, and every $\$$ substituted with the outer separator gadget $1^{d+1}$.

For any $b\in B$, the length of the string $p(b)$ is $(d+2)d$, since it contains $d+1$ inner separator gadgets of length $d$.
The string $P$ consists of $m+1$ outer separator gadgets, each of length $d+1$, and $m$ substrings $p(b)$ each of length $(d+2)d$. The length of $P$ is therefore $(d+2)dm+(m+1)(d+1)=\Theta(md^2)$.

For any $a\in A$, the length of the string $s(a)$  is $(d+3)d$.
Analogously to before, we define a \emph{block} as a substring of the form $s(a)\  1^{d+1}$. The length of a block is $(d+3)d+(d+1)=d^2+4d+1$.
The string $S$ consists of $4n+1$ blocks (as before), each of length $d^2+4d+1$ and an extra outer separator gadget at the beginning. The length of $S$ is therefore $(d^2 +4d +1)(4n+1) + d +1=\Theta(nd^2)$.
 
 Now, if we align every $p(b)$ to a copy of $s(z)$, as in the proof in Section~\ref{sec:sizefour}, we need $2m-1$ blocks. Thus, we get a substring of length $w= (d^2 +4d +1)(2m-1) + d +1$. Our reduction will again depend on the fact that if there are no orthogonal vectors, we cannot do much better than that. Namely, we next prove that if there is no pair of orthogonal vectors then there is no substring of $S$ of length $< w-2d$ that contains $P$ as a subsequence (Claim~\ref{claim:a}), and that if there is a pair of orthogonal vectors then there is a substring of $S$ of length $\leq w-(d^2 +4d +1)$ that contains $P$ as a subsequence (Claim~\ref{claim:b}). 

\medskip
\noindent
{\bf Algorithm.} We run the assumed blackbox algorithm for Episode Matching on $S$ and $P$. If the algorithm outputs a substring of length at least $w-2d$, we conclude that there are no orthogonal vectors. If it outputs some string of a shorter length, then there are.

\medskip
\noindent
{\bf Correctness.} The correctness proof follows along the same lines as in the alphabet four case. 
We begin by proving an equivalent version of Lemma~\ref{lem:subseq_four}:

\begin{lemma}
For two vectors $a$ and $b$ of dimension $d$, the string $p(b)$ is a subsequence of $s(a)$ if and only if $b$ and $a$ are orthogonal.
\end{lemma}
\begin{proof}
If $a$ and $b$ are orthogonal, then by the same arguments as before, $p(b)$ is a subsequence of $s(a)$. Otherwise, $a$ and $b$ are not orthogonal, so there exists an $i$ such that $b[i]=a[i]=1$. Assume for the sake of contradiction that $p(b)$ is a subsequence of $s(a)$. Then we must align the 1 corresponding to $b[i]$ with some coordinate gadget $s(a[j])=01$ and $j\neq i$. There are two cases depending on whether $i>j$ or $j>i$.  

 \begin{itemize}
        \item Case 1: $i>j$. Note that to the left of the 1 corresponding to $b[i]$, there are $(i-1)+di$ characters in $p(b)$, and to the left of the 1 in $s(a[j])=01$ there are $2(j-1)+dj+1$ characters in $s(a)$: namely $j-1$ coordinate gadgets, $j$ inner separator gadgets, and the $0$ in $s(a[j])$. But this means that the prefix of $p(b)$ up to $b[i]$ is longer than the prefix of $s(a)$ up to the 1 in $s(a[j])$, since
$$(i-1)+id \; \geq \; j+(j+1)d\; =\; j+d+jd\; >\; 2(j-1)+jd+1,$$
where the last inequality holds because $j< d$. Thus we cannot have aligned the prefix of $p(b)$ up to $b[i]$ to the prefix of $s(a)$ up to the 1 in $s(a[j])$, contradicting that $p(b)$ is a subsequence of $s(a)$.

 \item Case 2:    $j>i$. Note that to the right of the 1 corresponding to $b[i]$, there are $d-i+(d-i+1)d=(d-i)(d+1)+d$ characters in $p(b)$, and to the right of the 1 in $s(a[j])=01$ there are $2(d-j)+(d-j+1)d=(d-j)(d+2)+d$ characters in $s(a)$. But this means that the suffix of $p(b)$ to the right of $b[i]$ is longer than the suffix of $s(a)$ to the right of $s(a[j])$, since
    $$
        (d-i)(d+1)+d\geq(d-j+1)(d+1)+d=(d-j)(d+1)+2d+1>(d-j)(d+2)+d.
    $$
   This means that we cannot have aligned the suffix of $p(b)$ to the right of $b[i]$ to the suffix of $s(a)$ to the right of $s(a[j])$, contradicting that $p(b)$ is a subsequence of $s(a)$.\qedhere
   \end{itemize}
\end{proof}
Next we prove that if there are no vectors $a\in A$ and $b\in B$ which are orthogonal, then we cannot do much better than aligning $p(b_1), \dots, p(b_m)$ to $m$ consecutive copies of $s(z)$. Recall that we defined the length of that alignment as $w$.
 \begin{claim}\label{claim:a} If there exist no $a\in A$ and $b\in B$ which are orthogonal, then there exists no substring of $S$ of length $<w-2d$ that contains $P$ as a subsequence.\end{claim}
 
 \begin{proof}
    Assume an alignment such that $p(b_i)$ is not aligned to $s(z)$. Since there are no orthogonal vectors, there is no $s(a)$ such that $p(b_i)$ is aligned in $s(a)$. Further, since $p(b_i)$ starts and ends with a 0, its alignment cannot start or end within an outer separator gadget. Thus, the alignment of $p(b_i)$ spans a string either containing
    \begin{enumerate}
        \item \label{case1} a non-empty suffix of some $s(a_j)$, followed by $1^{d+1}$, followed by a non-empty prefix of $s(z)$  or
        \item \label{case2} a non-empty suffix of $s(z)$, followed by $1^{d+1}$, followed by a non-empty prefix of some $s(a_j)$.
    \end{enumerate}
    Consider case \ref{case1}. Since $s(z)$ only contains $d$ 1s, at least one character in the outer separator gadget following $p(b_i)$ cannot be aligned in $s(z)$. Thus, at least one 1 cannot be aligned before the $1^{d+1}$ following $s(z)$. If $i<m$, then the inner separator gadget at the beginning of $p(b_{i+1})$ cannot be aligned before the beginning of a new block. Thus, the alignment defined by aligning $p(b_i)~1^{d+1}$ to $s(z)~1^{d+1}$ spans a string no longer than the original alignment. If $i=m$, the same alignment spans a string at most $d$ longer than the original alignment. Analogously, for case \ref{case2}, we align $1^{d+1}~p(b_i)$ to $1^{d+1}~s(z)$. The alignment spans a string of the same length as the original alignment, or at most $d$ longer if $i=1$. Repeating this step for any $p(b_i)$ which is not aligned to some $s(z)$, we get an alignment where every $p(b_i)$ is aligned to some $s(z)$, and which spans a string at most $2d$ longer than the original alignment. This concludes the proof.
 \end{proof}

\begin{claim}\label{claim:b}
If there exist $a$ and $b$ which are orthogonal, then there is a substring of $S$ of length $\leq w-(d^2 +4d +1)$ (i.e. it is shorter than $w$ by at least a block's length) that contains $P$ as a subsequence.
\end{claim}
\begin{proof}
    The proof is analogous to that of Claim \ref{claim_short}.
\end{proof}

\noindent
{\bf Analysis.} The length of $S$ is now $\Theta(nd^2)$, and the length of $P$ is $\Theta(md^2)$. The contradictions to OVH and UOVH are constructed analogously to the alphabet four case. This proves Theorem \ref{thm:lower_bound_ovh} restricted to binary alphabet.

\section{Episode Matching Indexing }\label{sec:indexing}

In this section we consider the indexing version of episode matching: Given the string $S$ and an integer $k$, construct a data structure that can quickly report the episode matching of any pattern $P$ of length $k$.

\subsection{Upper bound}\label{subsec:tstradeoff}
We now prove Theorem~\ref{thm:ds_upper_bound}. 
We call letters that appear more than $\tau$ times \emph{frequent} letters. Note that there are at most $\frac{n}{\tau}$ frequent letters. For every $k$-tuple of frequent letters, we precompute and store the answer in a table. The size of this table is therefore $\left(\frac{n}{\tau}\right)^k$. 
Additionally, for each letter in the alphabet, we store a predecessor data structure containing all positions in $S$ where the letter appears.
Using a linear-space predecessor data structure such as y-fast-trie~\cite{Willard1983}, this requires an additional $O(n)$ space (every position in $S$ appears in exactly one predecessor structure) and answers predecessor/successor queries in $O(\log\log n)$ time. 

To answer a query, given a pattern $P$ of length $k$, if every letter in $P$ is frequent, we simply return the precomputed answer from the table. Otherwise, suppose $P[j]$ is a non-frequent letter. For each position $i$ s.t $S[i] = P[j]$, we find the minimal substring of $S$ that contains $P$ and aligns $P[j]$ to $S[i]$ (eventually we return the smallest substring found). To do this, we start from location $i$ and use successor queries to find $P[j+1],P[j+2],\ldots, P[m-1]$ and predecessor queries to find $P[j-1],P[j-2],\ldots, P[0]$. Since there are at most $\tau$ positions in $S$ that contain $P[j]$, this takes overall $O(\tau \cdot  k \log \log n)$ time.

\subsection{Lower bound}\label{sec:lowerindexing}
We now prove Theorem~\ref{thm:ds_lower_bound}. The proof is similar to the ones in \cite{DBLP:conf/cpm/KopelowitzK16} and \cite{DBLP:conf/cpm/BilleGPS21}. 

Recall that in the $k$-Set Disjointness problem, given sets $S_1, \ldots, S_m$ of total size $N$ over a universe $U$, we want a data structure that given $(i_1,\dots, i_k)$ reports whether $\bigcap_{j=1}^k S_{i_j}=\emptyset$.  As in \cite{DBLP:conf/cpm/BilleGPS21}, we can reduce $k$-Set Disjointness to $O(\log N)$ instances of $k$-Set Disjointness (each of size $O(N)$) with the property that every element in $U$ appears in the same number of sets (call this number $f$). We next show a reduction from such an instance to episode matching indexing.

Define for each set $S_i$ a unique letter $\alpha_i$. For each distinct element $e\in U$ define a block consisting of the letters $\alpha_i$ corresponding to the sets $S_i$ that contain $e$, sorted by $i$. We then append all such blocks in an arbitrary order and separate each two blocks by an extra block of the form $\$^f$ (where $\$$ is an extra letter not corresponding to any set). The resulting string $S$ is of length  $2N-f=O(N)$. We preprocess $S$ into a data structure for episode matching.
  To answer a query $(i_1,\dots,i_k)$ (we assume w.l.o.g that $i_1<i_2<\cdots <i_k$), we query the data structure for  $P=\alpha_{i_1}\alpha_{i_2}\cdots\alpha_{i_k}$. If the output is larger than  $f$, we answer that the sets are  disjoint, otherwise we answer that they are not. The correctness is simple: If there is an element $e$ that appears in all sets $S_{i_1},\dots, S_{i_k}$, then $e$'s block contains $\alpha_{i_1} \alpha_{i_2} \cdots \alpha_{i_k}$ (and since we sorted the elements, they will appear in the right order). If on the other hand the sets are disjoint, then there is no block containing all letters, so the smallest substring containing all letters must include at least one $\$^f$ block (and is thus longer than $f$).
    
     The total space for all the $O(\log N)$ instances is therefore $O(\log N\cdot s_{episode}(N))$ and the runtime is $O(\log N \cdot t_{episode}(N))$, where $s_{episode}(n)$ and $t_{episode}(n)$ are the space and query time for a data structure for episode Matching indexing. To prove Theorem \ref{thm:ds_lower_bound}, assume for contradiction that $t_{episode}(n)=O(n^{\delta})$ and $s_{episode}(n)=O(n^{k-k\delta-\epsilon})$. Then the space for solving $k$-Set Disjointness is $O(N^{k-k\delta-\epsilon}\log N)=O(N^{k-k\delta-\epsilon'})$, for any $0<\epsilon'<\epsilon$, and the time is $O(N^{\delta}\log N)=O(N^{\delta+\epsilon''})$, for arbitrarily small $\epsilon''>0$. Setting $\epsilon''<\epsilon'/k$, we obtain a contradiction to Conjecture \ref{con:kset}. 
     
\subsection{The special case of $k=2$}
\label{sec:kequals2}

We now prove Theorem~\ref{thm:minplus}. Recall that, given the string $S$, our goal is to compute the binary matrix $D$ where $D[a,b]=1$ if and only if $S$ contains a substring of length at most $t$ that starts with the letter $a$ and ends with the letter $b$. 
Naively, we can compute $D$ in time $\min\{\tilde{O}(n\sigma+\sigma^2),O(nt+\sigma^2)\}$. %First, initialize every entry of $D$ with zero. 
To get  $\tilde{O}(n\sigma+\sigma^2)$, we   construct a predecessor data structure  for each letter, containing the occurrences of this letter in $S$. Then,  for each position in $S$ we find the succeeding occurrence of each letter in the alphabet, using the corresponding predecessor data structure. Whenever we find a pair of letters $(a,b)$ at  distance at most $t$, we set $D[a,b]$ to 1. To get $O(nt+\sigma^2)$, for each position in $S$ we scan $t$ entries forward and update $D$ as we go. 
    In the remainder of this section, we will show how to compute $D$ faster using min-plus multiplication of bounded integer matrices:

    \begin{lemma}[\cite{DBLP:conf/soda/WilliamsX20}]\label{lem:min_plus_prod}
    Given two $n\times n$ matrices $A$ and $B$, where $A$ has entries in $\{-M,\dots, M\}\cup \{\infty\}$ and $B$ is arbitrary, we can compute their min-plus product $C=A\odot B$ $($defined as $C[i,j]=\min_{k} (A[i,k]+ B[k,j]))$ in $\tilde{O}(\sqrt{M}n^{(3+\omega)/2})$ time.
    \end{lemma}

First, we divide the string $S$ into blocks $B_1, B_2, \dots, B_{ n/t}$, each of length $t$ (except maybe the last). Let $\delta_{\mathrm{first}}(a,j)$ (resp. $\delta_{\mathrm{last}}(a,j)$) be the distance from the beginning (resp. end) of the $j$th block to the first (resp. last) $a$ in that block,  and $\infty$ if the block has no $a$. Note that $ab$ is a subsequence of a length $t$ substring of $S$ if and only if one of the following two conditions holds:\\

          \noindent {\bf Condition 1:} There is a $j$ such that $\delta_{\mathrm{first}}(a,j)+\delta_{\mathrm{last}}(b,j)\leq t$. 
            To check this condition, we check if $(M_1\odot M_2)[a,b]\leq t$ where $M_1$ is the $\sigma\times  n/t$ matrix with $M_1[a,j]=\delta_{\mathrm{first}}(a,j)$, and $M_2$ is the $  n/t\times\sigma$ matrix with $M_2[j,b]=\delta_{\mathrm{last}}(b,j)$. \\

          \noindent   {\bf Condition 2:} There is a $j$ such that $\delta_{\mathrm{last}}(a,j)+\delta_{\mathrm{first}}(b,j+1)\leq t$. 
            To check this condition, we check if $(M_3\odot M_4)[a,b]\leq t$ where $M_3$ is the $\sigma\times ( n/t-1)$ matrix with $M_3[a,j]=\delta_{\mathrm{last}}(a,j)$, and $M_4$ is the $ ( n/t -1)\times\sigma$ matrix with $M_4[j,b]=\delta_{\mathrm{first}}(b,j+1)$.\\
        
        By the above discussion, it only remains to compute the products   $M_1\odot M_2$ and $M_3\odot M_4$ (as then each entry of $D$ can be found in constant time). We focus on $M_1\odot M_2$ (the other is symmetric). Observe that the entries of $M_1$ and $M_2$  are integers bounded by $t$. Therefore, we can hope to use Lemma \ref{lem:min_plus_prod} (with $M=t$) to multiply them. However, in Lemma \ref{lem:min_plus_prod} the matrices are square while our matrices are rectangular. To deal with this, we break the matrices into rectangular submatrices according to the value of $\sigma$.

    \begin{itemize}
        \item If $\sigma< n/t $. 
       We split $M_1$ into $M_{1,1}\dots, M_{1,\frac{n}{t\sigma}}$, each consisting of $\sigma$ consecutive columns. 
       We split $M_2$ into $M_{2,1}\dots, M_{2,\frac{n}{t\sigma}}$, each consisting of $\sigma$ consecutive rows.
    Note that $(M_1\odot M_2)[a,b]=\min_{k=1,\dots, \frac{n}{t\sigma}}(M_{1,k}\odot M_{2,k})[a,b]$.
         We can thus compute $M_1\odot M_2$ by computing $\frac{n}{t\sigma}$ min-plus products of $\sigma\times\sigma$ matrices, and setting each entry in $M_1\odot M_2$ to be the minimum of the corresponding entries in the $\frac{n}{t\sigma}$ output matrices.
        Using Lemma \ref{lem:min_plus_prod}, this takes time $O(\frac{n}{t\sigma}(\sigma^2 + \sigma^{(3+\omega)/2}\sqrt{t}))=O(\sigma^{1/2+\omega/2}\left(\frac{n}{t}\right)^{1/2} \sqrt{n})$. 
   
        \item If $\sigma\geq  n/t $. We split $M_1$ into $M_{1,1}\dots, M_{1,\frac{t\sigma}{n}}$, each consisting of $n/t$ consecutive rows. We split $M_2$ into $M_{2,1}\dots, M_{2,\frac{t\sigma}{n}}$, each consisting of $n/t$ consecutive columns. We compute $M_1\odot M_2$ by computing  $\left(\frac{t\sigma}{n}\right)^2$ min-plus products of $n/t\times n/t$ matrices $M_{1,i}\odot M_{2,j}$ for every $i,j$.   
         Using Lemma \ref{lem:min_plus_prod}, this takes time 
         $O(\left(\frac{t\sigma}{n}\right)^2 \cdot \left(\frac{n}{t}\right)^{(3+\omega)/2}\sqrt{t})=O(\sigma^2\left(\frac{n}{t}\right)^{\omega/2-1}\sqrt{n})$. 
    \end{itemize}
    This proves Theorem~\ref{thm:minplus}.

\bibliographystyle{plainurl}
\bibliography{bibliography}

\begin{thebibliography}{10}

\bibitem{focs/AbboudBW15}
Amir Abboud, Arturs Backurs, and Virginia~Vassilevska Williams.
\newblock Tight hardness results for {LCS} and other sequence similarity measures.
\newblock In {\em Proc. 56th {FOCS}}, pages 59--78, 2015.

\bibitem{AbboudW21}
Amir Abboud and Virginia~Vassilevska Williams.
\newblock Fine-grained hardness for edit distance to a fixed sequence.
\newblock In {\em Proc. 48th {ICALP}}, volume 198, pages 7:1--7:14, 2021.

\bibitem{icalp/AbboudWW14}
Amir Abboud, Virginia~Vassilevska Williams, and Oren Weimann.
\newblock Consequences of faster alignment of sequences.
\newblock In {\em Proc. 41st {ICALP}}, pages 39--51, 2014.

\bibitem{dke/AcharAS13}
Avinash Achar, A.~Ibrahim, and P.~S. Sastry.
\newblock Pattern-growth based frequent serial episode discovery.
\newblock {\em Data Knowl. Eng.}, 87:91--108, 2013.

\bibitem{soda/AlmanW21}
Josh Alman and Virginia~Vassilevska Williams.
\newblock A refined laser method and faster matrix multiplication.
\newblock In D{\'{a}}niel Marx, editor, {\em Proc. 32nd {SODA}}, pages 522--539, 2021.

\bibitem{iandc/ApostolicoA02}
Alberto Apostolico and Mikhail~J. Atallah.
\newblock Compact recognizers of episode sequences.
\newblock {\em Inf. Comput.}, 174(2):180--192, 2002.

\bibitem{icdm/AtallahGS04}
Mikhail~J. Atallah, Robert Gwadera, and Wojciech Szpankowski.
\newblock Detection of significant sets of episodes in event sequences.
\newblock In {\em Proc. 4th {ICDM}}, pages 3--10, 2004.

\bibitem{focs/BackursI16}
Arturs Backurs and Piotr Indyk.
\newblock Which regular expression patterns are hard to match?
\newblock In {\em Proc. 57th {FOCS}}, pages 457--466, 2016.

\bibitem{jcomp/BackursI18}
Arturs Backurs and Piotr Indyk.
\newblock Edit distance cannot be computed in strongly subquadratic time (unless {SETH} is false).
\newblock {\em {SIAM} J. Comput.}, 47(3):1087--1097, 2018.

\bibitem{DBLP:conf/cpm/BilleGPS21}
Philip Bille, Inge~Li G{\o}rtz, Max~Rish{\o}j Pedersen, and Teresa~Anna Steiner.
\newblock Gapped indexing for consecutive occurrences.
\newblock In {\em Proc. 32nd {CPM}}, pages 10:1--10:19, 2021.

\bibitem{apal/BoassonCGM01}
Luc Boasson, Patrick C{\'{e}}gielski, Ir{\`{e}}ne Guessarian, and Yuri~V. Matiyasevich.
\newblock Window-accumulated subsequence matching problem is linear.
\newblock {\em Ann. Pure Appl. Log.}, 113(1-3):59--80, 2001.

\bibitem{focs/BringmannK15}
Karl Bringmann and Marvin K{\"{u}}nnemann.
\newblock Quadratic conditional lower bounds for string problems and dynamic time warping.
\newblock In {\em Proc. 56th {FOCS}}, pages 79--97, 2015.

\bibitem{ipl/CegielskiGM06}
Patrick C{\'{e}}gielski, Ir{\`{e}}ne Guessarian, and Yuri~V. Matiyasevich.
\newblock Multiple serial episodes matching.
\newblock {\em Inf. Process. Lett.}, 98(6):211--218, 2006.

\bibitem{njc/CrochemoreIMRTT02}
Maxime Crochemore, Costas~S. Iliopoulos, Christos Makris, Wojciech Rytter, Athanasios~K. Tsakalidis, and T.~Tsichlas.
\newblock Approximate string matching with gaps.
\newblock {\em Nord. J. Comput.}, 9(1):54--65, 2002.

\bibitem{cpm/DasFGGK97}
Gautam Das, Rudolf Fleischer, Leszek Gasieniec, Dimitrios Gunopulos, and Juha K{\"{a}}rkk{\"{a}}inen.
\newblock Episode matching.
\newblock In {\em Proc. 8th {CPM}}, pages 12--27, 1997.

\bibitem{algorithmica/DurajKP19}
Lech Duraj, Marvin K{\"{u}}nnemann, and Adam Polak.
\newblock Tight conditional lower bounds for longest common increasing subsequence.
\newblock {\em Algorithmica}, 81(10):3968--3992, 2019.

\bibitem{icalp/EquiGMT19}
Massimo Equi, Roberto Grossi, Veli M{\"{a}}kinen, and Alexandru~I. Tomescu.
\newblock On the complexity of string matching for graphs.
\newblock In {\em Proc. 46th {ICALP}}, pages 55:1--55:15, 2019.

\bibitem{EquiMT21}
Massimo Equi, Veli M{\"{a}}kinen, and Alexandru~I. Tomescu.
\newblock Graphs cannot be indexed in polynomial time for sub-quadratic time string matching, unless {SETH} fails.
\newblock In {\em Proc. 27th {SOFSEM}}, volume 12607, pages 608--622, 2021.

\bibitem{spire/Gibney20}
Daniel Gibney.
\newblock An efficient elastic-degenerate text index? not likely.
\newblock In {\em Proc. 27th {SPIRE}}, pages 76--88, 2020.

\bibitem{DBLP:conf/wads/GoldsteinKLP17}
Isaac Goldstein, Tsvi Kopelowitz, Moshe Lewenstein, and Ely Porat.
\newblock Conditional lower bounds for space/time tradeoffs.
\newblock In {\em Proc. 15th {WADS}}, pages 421--436, 2017.

\bibitem{sdm/GwaderaAS05}
Robert Gwadera, Mikhail~J. Atallah, and Wojciech Szpankowski.
\newblock Markov models for identification of significant episodes.
\newblock In {\em Proc. 5th {SDM}}, pages 404--414, 2005.

\bibitem{kais/GwaderaAS05}
Robert Gwadera, Mikhail~J. Atallah, and Wojciech Szpankowski.
\newblock Reliable detection of episodes in event sequences.
\newblock {\em Knowl. Inf. Syst.}, 7(4):415--437, 2005.

\bibitem{dis/HiraoISTA01}
Masahiro Hirao, Shunsuke Inenaga, Ayumi Shinohara, Masayuki Takeda, and Setsuo Arikawa.
\newblock A practical algorithm to find the best episode patterns.
\newblock In {\em Proc. 4th {DS}}, pages 435--440, 2001.

\bibitem{jacm/Hirschberg77}
Daniel~S. Hirschberg.
\newblock Algorithms for the longest common subsequence problem.
\newblock {\em J. {ACM}}, 24(4):664--675, 1977.

\bibitem{jcss/ImpagliazzoP01}
Russell Impagliazzo and Ramamohan Paturi.
\newblock On the complexity of k-sat.
\newblock {\em J. Comput. Syst. Sci.}, 62(2):367--375, 2001.

\bibitem{jnsm/KlemettinenMT99}
Mika Klemettinen, Heikki Mannila, and Hannu Toivonen.
\newblock Rule discovery in telecommunication alarm data.
\newblock {\em J. Netw. Syst. Manag.}, 7(4):395--423, 1999.

\bibitem{algorithmica/KociumakaRS19}
Tomasz Kociumaka, Jakub Radoszewski, and Tatiana Starikovskaya.
\newblock Longest common substring with approximately k mismatches.
\newblock {\em Algorithmica}, 81(6):2633--2652, 2019.

\bibitem{DBLP:conf/cpm/KopelowitzK16}
Tsvi Kopelowitz and Robert Krauthgamer.
\newblock Color-distance oracles and snippets.
\newblock In {\em Proc. 27th {CPM}}, pages 24:1--24:10, 2016.

\bibitem{fuin/KuriNM03}
Josu{\'{e}} Kuri, Gonzalo Navarro, and Ludovic M{\'{e}}.
\newblock Fast multipattern search algorithms for intrusion detection.
\newblock {\em Fundam. Informaticae}, 56(1-2):23--49, 2003.

\bibitem{jal/MakinenNU05}
Veli M{\"{a}}kinen, Gonzalo Navarro, and Esko Ukkonen.
\newblock Transposition invariant string matching.
\newblock {\em J. Algorithms}, 56(2):124--153, 2005.

\bibitem{datamine/MannilaTV97}
Heikki Mannila, Hannu Toivonen, and A.~Inkeri Verkamo.
\newblock Discovery of frequent episodes in event sequences.
\newblock {\em Data Min. Knowl. Discov.}, 1(3):259--289, 1997.

\bibitem{nowickacomplexity}
El{\.z}bieta Nowicka and Marcin Zawada.
\newblock On the complexity of matching non-injective general episodes.
\newblock {\em Computation and Logic in the Real World}, pages 288--296, 2007.

\bibitem{ipl/Polak18}
Adam Polak.
\newblock Why is it hard to beat \emph{O}(\emph{n}\({}^{\mbox{2}}\)) for longest common weakly increasing subsequence?
\newblock {\em Inf. Process. Lett.}, 132:1--5, 2018.

\bibitem{dasfaa/TagoAKMI12}
Shinichiro Tago, Tatsuya Asai, Takashi Katoh, Hiroaki Morikawa, and Hiroya Inakoshi.
\newblock {EVIS:} {A} fast and scalable episode matching engine for massively parallel data streams.
\newblock In {\em Proc. 17th {DASFAA}}, pages 213--223, 2012.

\bibitem{Willard1983}
D.~E. Willard.
\newblock Log-logarithmic worst-case range queries are possible in space $\theta({N})$.
\newblock {\em Inf. Process. Lett.}, 17(2):81--84, 1983.

\bibitem{tcs/Williams05}
Ryan Williams.
\newblock A new algorithm for optimal 2-constraint satisfaction and its implications.
\newblock {\em Theor. Comput. Sci.}, 348(2-3):357--365, 2005.

\bibitem{DBLP:conf/soda/WilliamsX20}
Virginia~Vassilevska Williams and Yinzhan Xu.
\newblock Truly subcubic min-plus product for less structured matrices, with applications.
\newblock In {\em Proc.31st {SODA}}, pages 12--29, 2020.

\end{thebibliography}

\end{document}